\documentclass[conference,a4paper]{APSIPA2021}
\usepackage{multirow}
\usepackage{graphicx}
\usepackage{amsmath}
\usepackage[psamsfonts]{amssymb}
\usepackage{amsxtra}
\usepackage{threeparttable}
\usepackage{booktabs}
\usepackage{amssymb}
\usepackage{pifont}
\usepackage{cite}
\usepackage{subfigure}
\usepackage{color}

\newcommand{\cmark}{\ding{51}}%
\newcommand{\xmark}{\ding{55}}%

\begin{document}

\title{How Speech is Recognized to Be Emotional - A Study Based on Information Decomposition}

\author{%
\authorblockN{%
Haoran Sun, Lantian Li$^*$, Thomas Fang Zheng, Dong Wang$^*$
}
\authorblockA{%
Center for Speech and Language Technologies, BNRist, Tsinghua University \\
E-mail: lilt@cslt.org; wangdong99@mails.tsinghua.edu.cn}
}

\maketitle
\thispagestyle{empty}

\begin{abstract}

The way that humans encode their emotion into speech signals is complex.
For instance, an angry man may increase his pitch and speaking rate, and use impolite words.
In this paper, we present a preliminary study on various emotional factors and investigate how each of them impacts modern emotion recognition systems.
The key tool of our study is the SpeechFlow model presented recently,
by which we are able to decompose speech signals into separate information factors (content, pitch, rhythm).
Based on this decomposition, we carefully studied the performance of each information component and their combinations. 
We conducted the study on three different speech emotion corpora and chose an attention-based convolutional RNN as the emotion classifier.
Our results show that rhythm is the most important component for emotional expression.
Moreover, the cross-corpus results are very bad (even worse than guess),
demonstrating that the present speech emotion recognition model is rather weak.
Interestingly, by removing one or several unimportant components, the cross-corpus results can be improved.
This demonstrates the potential of the decomposition approach towards a generalizable emotion recognition.

\end{abstract}

\section{Introduction}

Recognizing emotion from speech signals is highly desirable for designing a comfortable human-machine interface.
After three decades of research, speech emotion recognition (SER) has gained significant improvement~\cite{schuller2018speech}. Early research
mostly focused on extracting emotion-related features, forming some `standard' feature sets such as GeMAPS~\cite{eyben2015geneva} and COMPARE\cite{schuller2013interspeech}.
By these emotion features, simple classifiers such as hidden Markov model (HMM) or support vector machines (SVM) were
employed to determine the emotion~\cite{schuller2011recognising,el2011survey} .
Recently, deep learning methods gained much popularity, in particular the end-to-end architecture based on 
deep neural nets (DNN)~\cite{satt2017efficient,kim2017deep,trigeorgis2016adieu,neumann2017attentive,chen20183,cheng2018mmann,kwon2021mlt,latif2019direct}.
Good performance was reported with various types of DNN models, including convolutional neural network (CNN)~\cite{mao2014learning}, deep belief network~\cite{wen2017random},
long-short term memory (LSTM)~\cite{trigeorgis2016adieu} and variational auto-encoders (VAE)~\cite{latif2017variational}. The main advantage of deep learning models
is that they can learn emotional cues automatically, so potentially discover features more powerful than human engineering.

\begin{figure*}[bpt]
\centering
\includegraphics[width=1.0\linewidth]{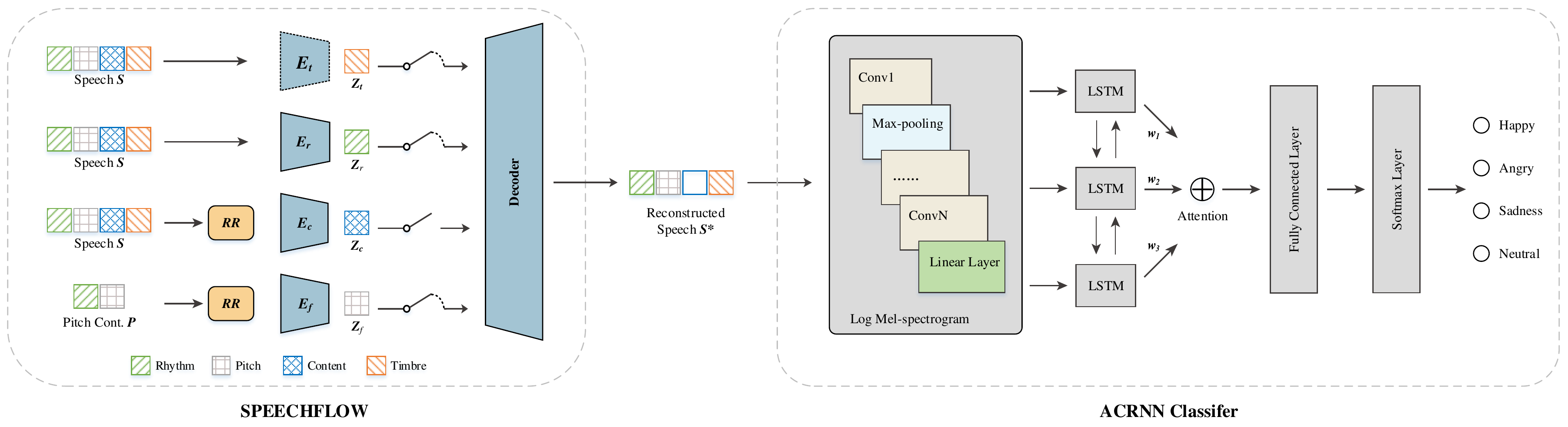}
\caption{The full diagram of the work. The SpeechFlow model decomposes speech signals into three information factors: content $Z_c$, rhythm $Z_r$, pitch $Z_f$.
These information factors can be manipulated in order to modify information load of the reconstructed speech signal, for instance set the same pitch for all the frames of
all the utterances. Note that an additional factor related to speaker trait (timbre) is also represented in the latent space and is denoted by $Z_t$, but it is
only used for speech reconstruction and is not modified in our study. Once the speech signal reconstructed from the modified factors, ACRNN-based SER model is used to predict
the emotion of the reconstructed speech.}
\label{fig:arch}
\end{figure*}

In spite of the notable advance in performance, how a speech is recognized by the machine to be emotional is still far from clear.
One reason is that human emotions in audio data are very complex, and the expression and perception of a particular emotion is impacted by
various factors such as gender, speakers, age, culture, and languages~\cite{kwon2021att}.
From the signal processing perspective, Murray et al~\cite{murray1993toward} identified that the quality of voice, the timing of pronunciation units,
and the pitch contour are mostly affected in emotional speech. This study is very inspiring and guided the long-standing research on emotion sensitive features.
However, it is still not easy to identify by which information factors in the speech signal that \emph{machines} recognize human's emotion, even if we
can test and compare the SER performance with individual features and their combinations. This is because there
is no guarantee that these features faithfully reflect the underlying information factors (e.g., prosody patterns), and there is no guarantee that the complex temporal/frequency dependencies
within the original speech signal can be recovered by reintegrating the separately extracted features.
This is in particular true with the DNN-based end-to-end model, as the decision is made largely in a black box.

In this paper, we try to answer the following question: ``How an end-to-end DNN model determines emotions''. Our main tool is a
speech factorization model called \emph{SpeechFlow}~\cite{qian2020unsupervised}. By this model, speech signals can be decomposed into separate information
factors, and these factors can be put together to recover the original speech.
This analysis-and-synthesis tool offers us an interesting opportunity to
manipulate the information load in speech signals, allowing us testing the impact of each individual information factor and their combinations.
In this preliminary study, we decompose speech signal into three
components: content, rhythm, and pitch. This decomposition was motivated by the importance of timing (rhythm) and pitch in human's emotion perception, as
found by Murray~\cite{murray1993toward}, as well as the intrinsic association of emotion status and linguistic content~\cite{grandjean2006intonation,pell2011emotional,mittal2020m3er}. 
Fig.~\ref{fig:arch} illustrates the full diagram of our approach.

We chose the attention-based convolutional RNN (ACRNN)~\cite{chen20183} as the SER backbone, due to its good performance reported in the literature.
The ACRNN model was trained with IEMOCAP, a popular
emotion speech dataset in English, and was tested on IEMOCAP, SAVEE and CSLT-ESDB, where SAVEE and CSLT-ESDB are two datasets in English and Chinese respectively.
The results show that among the three information factors (content, rhythm, pitch), rhythm is the most discriminative and generalizable.
When the test data is highly mismatched with the training data, for example in the cross-lingual test, removing some information factors may increase rather than decrease the performance.
This suggests that to achieve reasonable generalization, information control deserves deeper investigation.

Our contribution is two-fold: (1) We employed speech factorization as a novel tool to investigate the decision behavior of DNN, which is new in SER; (2)
Using the factorization tool, we studied the salient information factors in SER, and investigated the generalization capacity of different factors in cross-domain and cross-lingual
situations. The paper is organized as follows: in Section~\ref{sec:method}, we first briefly introduce the SpeechFlow and ACRNN models, and present our implementation details. 
Related work is discussed in Section~\ref{sec:rel}, and the experiments
are presented in Section~\ref{sec:exp}. Finally the paper is concluded by Section~\ref{sec:con}.

\section{Methods}
\label{sec:method}

In this section, we firstly revisit the SpeechFlow model presented in~\cite{qian2020unsupervised}. This model can decompose speech signals into 
separate information factors, and then reconstruct the original speech from these factors. Importantly, we can \emph{remove}
any individual factor by setting its value to be a constant, which allows us freely manipulating the information load of the speech signal.
The second component of our architecture is an attention-based convolutional RNN (ACRNN)~\cite{chen20183}. 
This model can use as the backbone of our SER system, and test the SER performance of speech signals with different information
load. The entire architecture of this work is illustrated in Fig.\ref{fig:arch}.

\subsection{SpeechFlow for speech information decomposition}

SpeechFlow is a speech factorization model proposed recently~\cite{qian2020unsupervised}. 
This model can decompose speech signals into separate information factors in an unsupervised way.
As shown in Fig.\ref{fig:arch} (left), speech $\emph{S}$ is decomposed by SpeechFlow into four information factors: rhythm $\textbf{Z}_r$, pitch $\textbf{Z}_f$, content $\textbf{Z}_c$ and timbre $\textbf{Z}_t$,
by four neural encoders: a \emph{rhythm encoder $\textbf{E}_r$}, a \emph{pitch encoder $\textbf{E}_f$}, a \emph{content encoder $\textbf{E}_c$} and a \emph{timbre encoder $\textbf{E}_t$}.

The training objective of the model is to reconstruct the original speech $\emph{S}$ from these information factors, with a \emph{decoder D}. 
Formally, the reconstructed speech $\hat{\emph{S}}$ is obtained by:

\begin{equation}
 \hat{\emph{S}} = \emph{D}(\emph{\textbf{Z}}_r, \emph{\textbf{Z}}_f, \emph{\textbf{Z}}_c, \emph{\textbf{Z}}_u), 
\end{equation}

\noindent and the objective function $\mathcal{L}$ is formulated as:

\begin{equation}
 \mathcal{L} = ||~\hat{\emph{S}} - \emph{S}~||^2.  
\end{equation}

\noindent where $||\cdot||$ denotes the $\ell_2$-norm.
To ensure that the information factors posses their own desired information after model training,
the encoders need some special designs, as shown below.

Firstly, the input to the rhythm encoder $\textbf{E}_r$, content encoder $\textbf{E}_c$ and timbre encoder $\textbf{E}_t$ is speech $\emph{S}$,
whereas the input to pitch encoder $\textbf{E}_f$ is the normalized pitch contour $\emph{P}$.
Note that $\emph{P}$ has been normalized to posses the same mean and variation for all the speakers, 
so it involves only rhythm and pitch information, without information about speaker trait and linguistic content.

Secondly, a random resampling operation ($\emph{\textbf{RR}}$) along the temporal axis is conducted before speech $\textbf{S}$ is fed to the content encoder $\textbf{E}_c$ and 
pitch encoder $\textbf{E}_f$. This operation randomly shrinks or stretches the duration of each speech segment. This operation makes the two encoders lose the true rhythm information, 
so the complete rhythm information can only pass through the rhythm encoder $\textbf{E}_r$ and represented by $\textbf{Z}_r$. 

Thirdly, the speaker identity vector is used as the timbre information. We choose a pre-trained speaker recognition model as the timbre encoder $\textbf{E}_t$. This model 
is based on deep speaker embedding~\cite{variani2014deep,li2019gaussian}, and the produced speaker vectors are used as $\textbf{Z}_t$. Note that $\textbf{Z}_t$ involves only the timber information.

Finally, the dimensions of all the information factors are much lower than the input speech. This limited dimensionality forms the so-called \emph{information bottleneck}, 
which means that none of the individual factors can represent the whole signal, and all the factors must cooperate together to achieve the learning goal, i.e., reconstruct the original speech. 
To make this cooperation effective and economic, each factor has to focus on the information that it can easily supply and others cannot, thus leading to the desired 
information decomposition~\cite{qian2020unsupervised}.

Put them together, the entire encoding process can be formulated as follows:

\begin{equation}
 \begin{aligned}
  &\emph{\textbf{Z}}_r = \emph{\textbf{E}}_r(\emph{S}),              \nonumber \\
  &\emph{\textbf{Z}}_f = \emph{\textbf{E}}_f(\emph{\textbf{RR}}(\emph{P})),   \nonumber \\
  &\emph{\textbf{Z}}_c = \emph{\textbf{E}}_c(\emph{\textbf{RR}}(\emph{S})),   \nonumber \\
  &\emph{\textbf{Z}}_t = \emph{\textbf{E}}_t(\emph{S}).              \nonumber
 \end{aligned}
\end{equation}

\noindent where $\emph{\textbf{RR}}$ denotes random resampling.

SpeechFlow offers a powerful analysis-and-synthesis tool, by which we can freely manipulate the information factors, hence modifying the information load in the reconstructed speech.
For example, we can set any information factor to be a constant, so that remove the corresponding information from the reconstructed speech. In this study, we focus on the impact of 
three information factors: content ($\textbf{Z}_c$), rhythm ($\textbf{Z}_r$), and pitch ($\textbf{Z}_f$). The impact of the timbre factor $\textbf{Z}_t$ will be left for future investigation.

\subsection{ACRNN-based emotion recognition}

If the SpeechFlow model has been well-trained, we can use it to manipulate the information load in speech signals. This allows us to 
conduct an ablation study to evaluate which information factor is the most important for emotion recognition. We first generate speech 
signals with particular information involved, and then fed the speech into an attention-based convolutional RNN (ACRNN)~\cite{chen20183} 
classifier for emotion recognition, as shown in Fig.\ref{fig:arch} (right). 

The ACRNN structure consists of four components, as detailed below. More details about ACRNN can be found in~\cite{chen20183}.

\begin{itemize}
\item \textbf{CNN component:} This component involves a 3-D convolutional layer followed by a max-pooling layer, upon which 
5 convolutional layers and 1 full-connection layer are stacked. This component aims to learn local patterns of emotion traits.

\item \textbf{RNN component:} It is a single-layer RNN with LSTM units. This component is designed to model the long-term patterns of emotion traits.

\item \textbf{Attention component:} This component aggregates the frame-level representations along the temporal axis, to form an utterance-level emotion representation.
In particular, the attention mechanism weights each frame according to the information that contains related to emotion prediction. 

\item \textbf{Prediction component:} It involves a full-connection layer and a softmax layer, and the output units of the softmax correspond to the emotion classes.
\end{itemize}

Note that for each configuration of the information factor selection in SpeechFlow, we need retrain the ACRNN model. This is the case even we 
select \emph{all} the information factors (i.e., do not intentionally remove any information).

\section{Related work}
\label{sec:rel}

Speech factorization has been extensively used in speech coding, speech synthesis, and voice conversion ~\cite{hsu2017learning,chou2018multi,zhang2019learning,zhou2021vaw},
but the application in SER is rare. Li et al.~\cite{li2018deep} proposed a cascade factorization approach, where content and speaker traits are sequentially extracted
in prior, and these information factors are used as conditional inputs of an end-to-end SER system.
Peri et al.~\cite{peri2021disentanglement} employed an adversarial training to purify emotion embeddings, by using both audio and visual streams, though it is more a feature extraction
approach rather than a complete factorization approach.

The research on the decision process of DNN in SER is also not extensive.
Jalal et al.~\cite{Jalal2020} investigated how an attention-based DNN model focuses on the important segment of the speech signal when performing SER. This work is
different from ours as we focus on important information factors rather than important segments.

\section{Experiments}
\label{sec:exp}

\subsection{Data}

The corpora used in our experiments are summarized in Table~\ref{tab:data}. Considering that the sets of emotion labels are not completely the same 
for different corpora, we chose to use the four overlapped emotion classes: \emph{\textbf{A}ngry}, \emph{\textbf{H}appy}, \emph{\textbf{S}ad} and \emph{\textbf{N}eutral}.
All the speech signals were uniformly formatted to 16kHz, 16-bits to ensure data consistency.

Since English and Chinese are different in pronunciation, we trained separate SpeechFlow models for the two languages, using VCTK and AISHELL-3 respectively.
Two ACRNN SER models were trained, one with IEMOCAP and the other with CSLT-ESDB\footnote{http://data.cslt.org}. For each training, 
the entire corpus was split into a training set (80\%), a validation set (10\%) and a test set (10\%).
The speakers and utterances in the validation and test sets did not appear in the training set. The SAVEE dataset is too small to be used for model training, we therefore
used it as a test set only.

\begin{table*}[]
 \caption{Corpora description}
 \centering
 \begin{tabular}{llllllll}
  \toprule
   \textbf{Corpus} & \textbf{Language} & \textbf{Content} &  \textbf{Emotion Types}  &  \textbf{\# Utters} & \textbf{\# Spks} & \textbf{\# Hours} & \textbf{Usage} \\
   \midrule
    VCTK~\cite{veaux2017cstr}           & English  &  Newspaper      &  -   &  5,376  &  20  &  5    &  SpeechFlow training \\
    AISHELL-3~\cite{AISHELL-3_2020}     & Chinese  &  Natural Text   &  -   &  6,162  &  20  &  6    &  SpeechFlow training \\
   \midrule
    IEMOCAP~\cite{busso2008iemocap}     & English  &  Dialogue       &  A, H, S, N   &  2,280  &  10  &  2.5  &  ACRNN training \& Emotion evaluation \\
    CSLT-ESDB~\cite{bie2013emotional}   & Chinese  &  Emotional Text &  A, H, S, N   &  7,200  &  30  &  10   &  ACRNN training \& Emotion evaluation \\
    SAVEE~\cite{wang2010machine}        & English  &  Natural Text   &  A, H, S, N   &  300    &  4   &  0.5  &  Emotion evaluation \\
   \bottomrule
  \label{tab:data}
 \end{tabular}
\end{table*}

\subsection{Configurations}

\subsubsection{SpeechFlow}

The SpeechFlow model was implemented using the source code published online\footnote{https://github.com/FantSun/Speechflow.}.
We mostly followed the settings in the original repository, including network structures, data preprocessing steps, and the training scheme.
The only change we made is the timbre encoder, which is one-hot speaker codes. This one-hot code 
cannot be extended to represent speakers outside of the training dataset, therefore not suited for our task. 

To solve this problem, we introduced a speaker encoder to realize the function of the timbre encoder. The speaker encoder is able to generate 
continuous vectors to represent the trait of speakers. These continuous vectors, often called \emph{speaker vectors}, are generalizable to 
speakers in any database. 
In our experiments, we chose the d-vector model to implement the speaker (timbre) encoder~\cite{variani2014deep,li2019gaussian}. The model was constructed 
following the Kaldi SITW recipe~\cite{povey2011kaldi}.

\subsubsection{ACRNN}

We built the ACRNN SER model using the public source code online\footnote{https://github.com/xuanjihe/speech-emotion-recognition.}.
The input feature was 80-dimensional mel-spectrograms, to match the output of the SpeechFlow model.
Other configurations were the same as in~\cite{chen20183}.

\subsection{Qualitative Analysis of SpeechFlow}

\begin{figure}[!htp]
\centering
\includegraphics[width=1.0\linewidth]{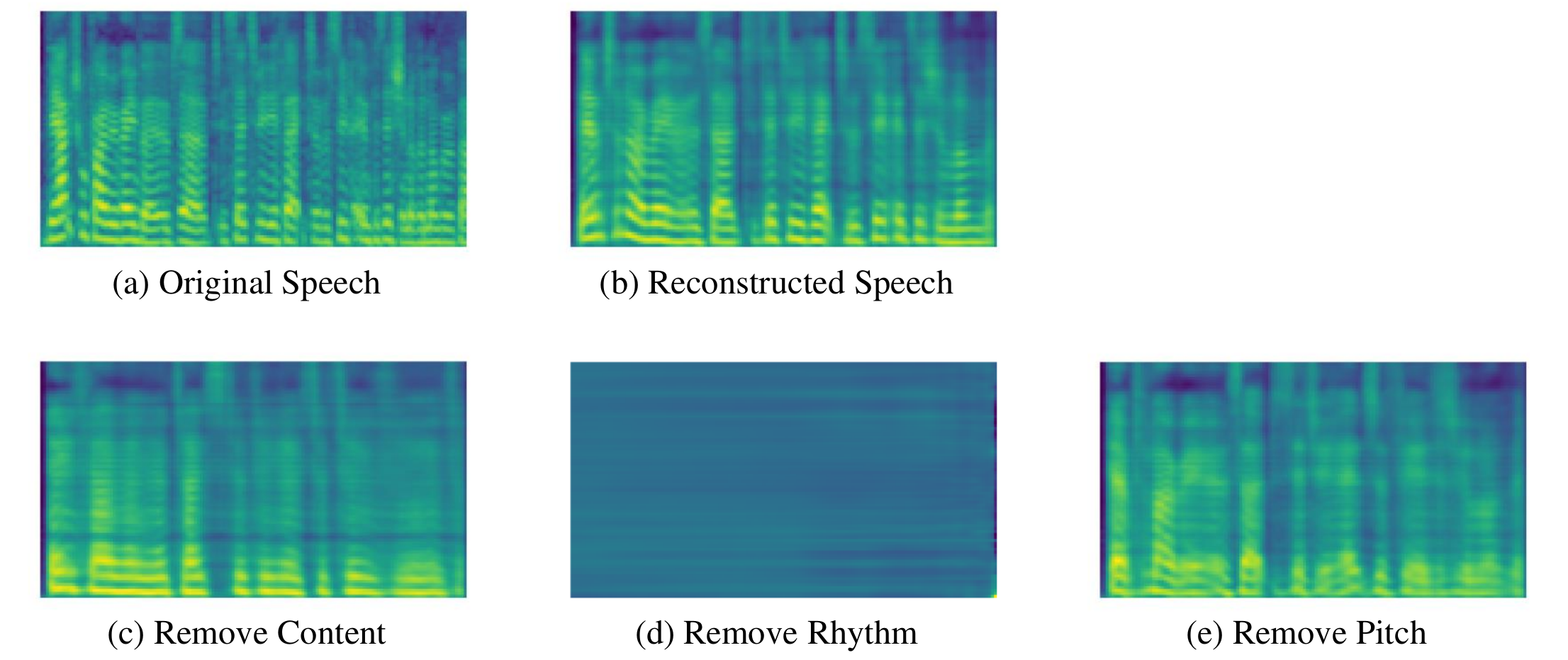}
\caption{The spectrogram reconstructed by SpeechFlow. The model was trained with VCTK, and the speech signal was selected from IEMOCAP.
(a) spectrogram of the original speech; (b) spectrogram of the reconstructed speech with all the information factors preserved; (c)$\sim$(e) spectrograms of
the reconstructed speech with a single information factor removed.}
\label{fig:ep}
\end{figure}

We firstly verify how SpeechFlow decomposes and recovers speech signals.
Specifically, we first decompose the spectrogram of a speech signal into individual information factors, and then remove a particular factor by setting the input of the corresponding encoder to be zero. 
Finally, the spectrogram can be recovered by the decoder of the SpeechFlow model.
The reconstructed spectrograms, when different information factors are removed, are shown in Fig.\ref{fig:ep}.

Firstly, by the comparison of Fig.\ref{fig:ep} (a) and Fig.\ref{fig:ep} (b), one can observe that the reconstructed spectrogram with all information factors remained 
is similar to the original spectrogram, and detailed local patterns are retained.
It indicates that the SpeechFlow model can successfully reconstruct the speech signal to a large extent. This is the foundation of all the following experiments. 
Secondly, removing any information factor leads to significant change in the reconstructed spectrogram, and the change is mutually different when different factors are removed.


\subsection{Basic results}

In this section, we report the SER performance when different information factors are removed. As the first step, 
we chose IEMOCAP as the training data to build the ACRNN model, and tested the performance on IEMOCAP and SAVEE. The SpeechFlow
model was trained using VCTK. We use the unweighted average recall (UAR) as the metric, and the results are shown in Table~\ref{tab:basic}.
Note that `\cmark' denotes that this factor was preserved while ``\xmark'' denotes that this factor was removed.
For example, in system 4, the ACRNN model was trained with speech reconstructed by preserving the content factor $\textbf{Z}_c$ only,
and the validation and test data were processed in the same way.

\begin{table}[!htp]
 \caption{UAR(\%) results of ACRNN model trained on \emph{IEMOCAP} with different information factor combinations.}
 \centering
 \begin{tabular}{lccccc}
   \toprule
   \multirow{2}{*}{No.} & \multicolumn{3}{c}{Factors} & \multicolumn{2}{c}{Test Sets}  \\
   \cmidrule(r){2-4}  \cmidrule(r){5-6}
       & \multicolumn{1}{c}{Content} & \multicolumn{1}{c}{Rhythm} & \multicolumn{1}{c}{Pitch} & \multicolumn{1}{c}{IEMOCAP} & \multicolumn{1}{c}{SAVEE} \\
   \cmidrule(r){1-1}     \cmidrule(r){2-4}  \cmidrule(r){5-6}
   1  & -        & -        & -        & 59.08  & 45.00  \\
   2  & \cmark   & \cmark   & \cmark   & 58.46  & 42.71  \\
   3  & \xmark   & \xmark   & \xmark   & 25.00  & 25.00  \\
   \cmidrule(r){1-1}  \cmidrule(r){2-6}
   4  & \cmark   & \xmark   & \xmark   & 43.27  & 30.21  \\
   5  & \xmark   & \cmark   & \xmark   & 57.85  & 39.79  \\
   6  & \xmark   & \xmark   & \cmark   & 41.74  & 24.79  \\
   \cmidrule(r){1-1}  \cmidrule(r){2-6}
   7  & \cmark   & \cmark   & \xmark   & 57.14  & 42.50  \\
   8  & \cmark   & \xmark   & \cmark   & 40.20  & 31.25  \\
   9  & \xmark   & \cmark   & \cmark   & 57.89  & 37.50  \\
   \bottomrule
\label{tab:basic}
\end{tabular}
\end{table}

Firstly, we observe that system 2, where all the information factors are preserved, can obtain performance comparable to system 1 on both the two test sets.
Besides, if all the information factors are removed (system 3), 
the decision is completely random.\footnote{Note that there are 4 emotion classes in total so the performance by random guess is 25\%.}
These results double confirmed that SpeechFlow can decompose speech signals into information factors
and these factors can be put together to recover the original speech.

Secondly, we find that rhythm obtains better performance than content and pitch (compare system 5 vs. 4 and 6).
This observation is consistent in both the within-corpus test (IEMOCAP) and the cross-corpus test (SAVEE).
It indicates that rhythm is a salient feature used by the ACRNN model: it is not only discriminative, but also generalizable. 

Thirdly, we observe that combining rhythm with other factors did not lead to clear advantage, and sometimes may lead to performance loss. 
This suggests that DNNs may rely on one or a few information to perform the decision, rather than the full information set. 
This behavior, however, may be related to the limited training data,
and deeper investigation with a larger dataset will give a more convincing conclusion.

\begin{figure}[!htp]
\centering
\includegraphics[width=1.0\linewidth]{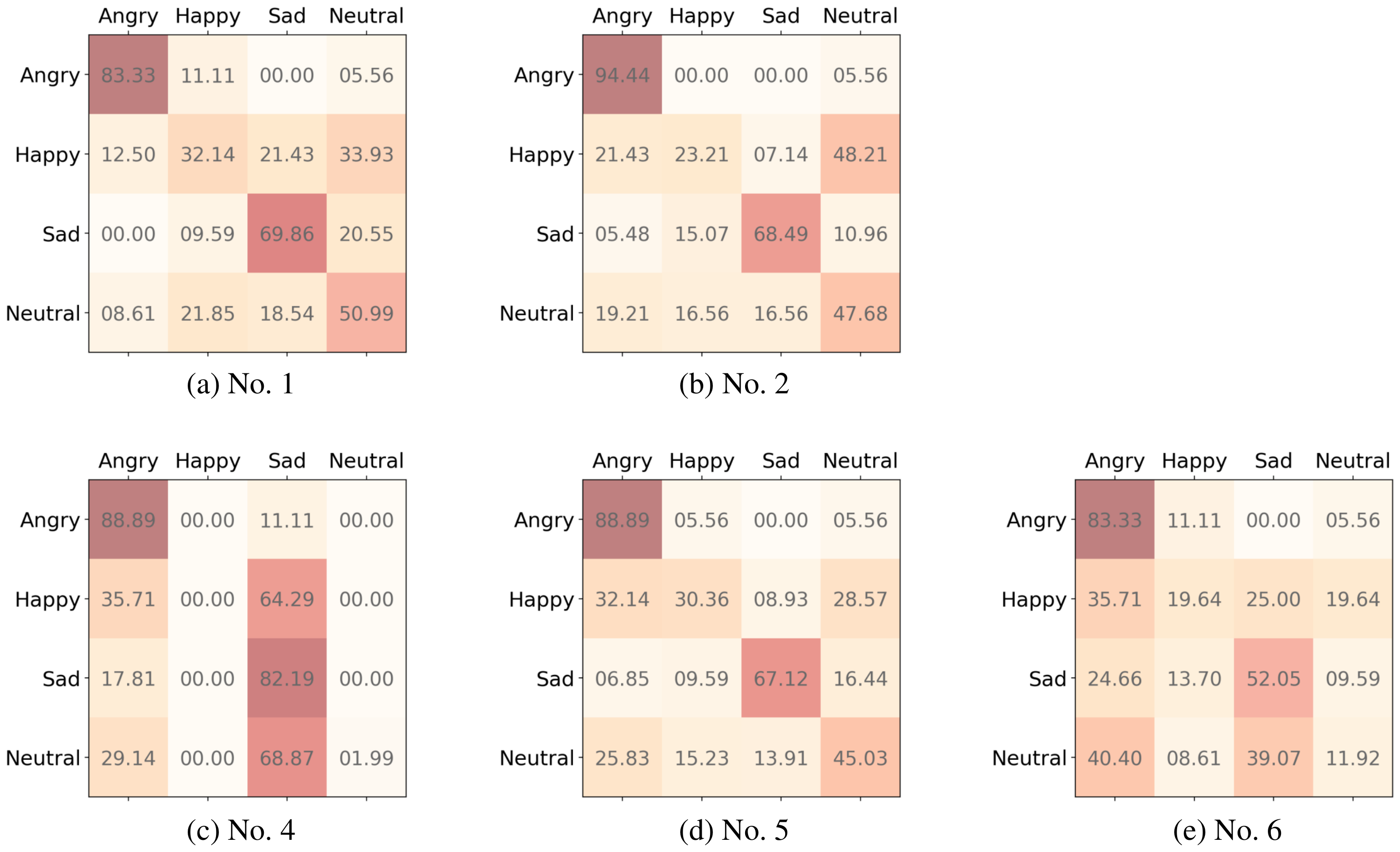}
\caption{Confusion matrices of systems in Table~\ref{tab:basic} when tested on IEMOCAP. The numbers in the 
cells represent the percentage that a ground-truth emotion (row label) is recognized as a particular emotion (column label).}
\label{fig:iem}
\end{figure}

\begin{figure}[!htp]
\centering
\includegraphics[width=1.0\linewidth]{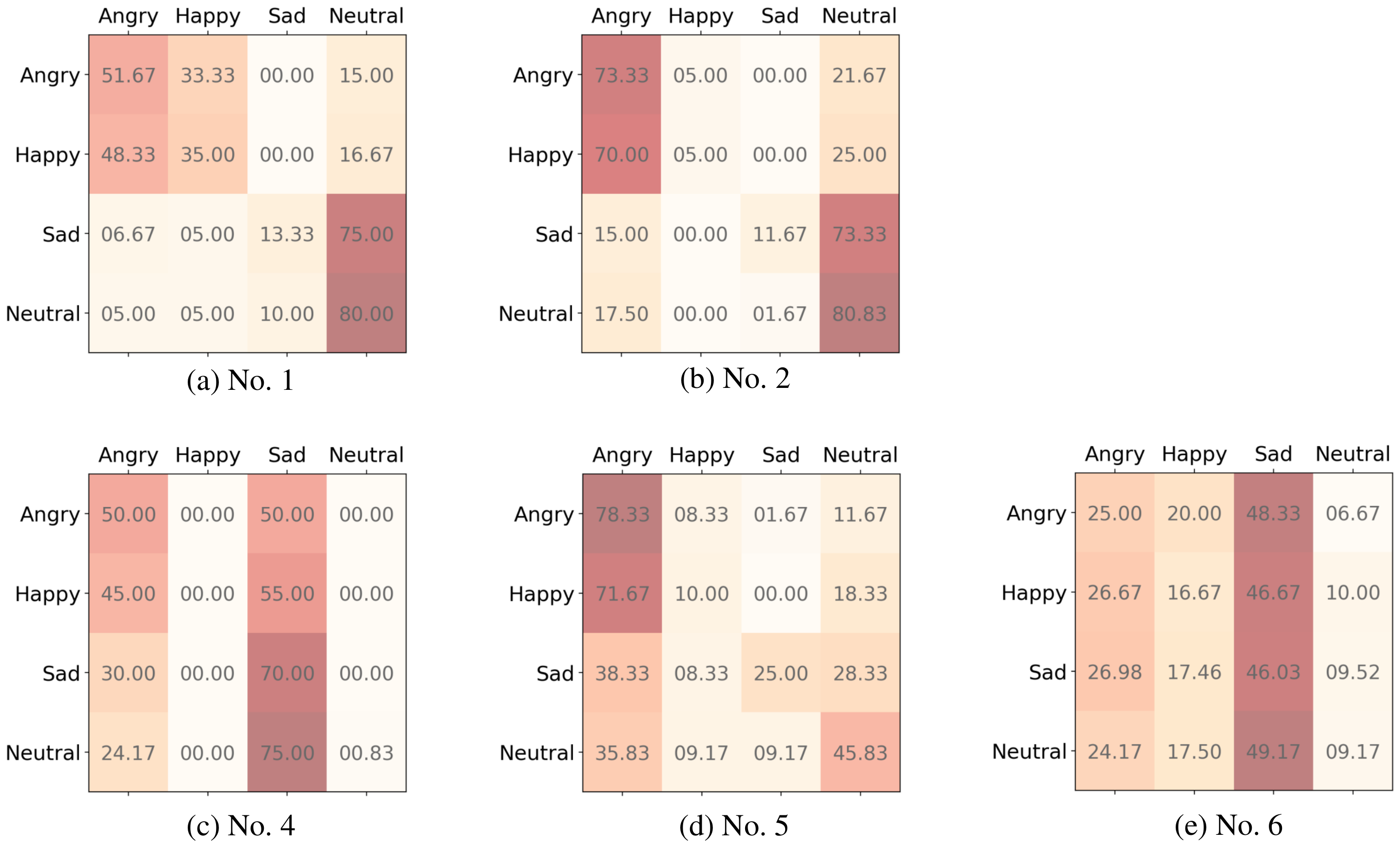}
\caption{Confusion matrices of systems in Table~\ref{tab:basic} when tested on SAVEE. The meaning of the labels and numbers are the same as in 
Fig.~\ref{fig:iem}.}
\label{fig:savee}
\end{figure}

To assist the analysis for the SER performances, the confusion matrices for
the IEMOCAP test and the SAVEE test are shown in Fig.\ref{fig:iem} and Fig.\ref{fig:savee}, respectively.
It can be found that the performance tendency is quite similar on the two test sets.
For example, for system 4 (content only), all the utterances tend to be recognized as \emph{sad}.
Moreover, the inter-emotion confusion is more balanced with system 5 (rhythm only) compared to system 4 (content only) and system 6 (pitch only),
double confirming the superiority of rhythm information in DNN-based SER.

\subsection{Cross-corpus results}
\label{sub:cross}

In this section, we design a cross-lingual test using the IEMOCAP and CSLT-ESDB datasets, which are in different languages.
VCTK and AISHELL-3 were firstly employed to train an English SpeechFlow and a Chinese SpeechFlow respectively.
Then, IEMOCAP was used to train the English ACRNN model, where the VCTK-based SpeechFlow was used to perform information selection, and CSLT-ESDB was 
used to train the Chinese ACRNN models, where the AISHELL-3-based SpeechFlow was used for information selection.
The UAR results are reported in Table~\ref{tab:cross}, where \emph{IEMO} and \emph{ESDB} are the abbreviations of IEMOCAP and CSLT-ESDB, respectively.

\begin{table}[!htp]
 \caption{UAR(\%) results with different information factor combinations on cross-lingual emotion recognition.}
 \centering
 \label{tab:cross}
 \scalebox{0.94}{
 \begin{tabular}{lccccccc}
   \toprule
   \multirow{2}{*}{No.} & \multicolumn{3}{c}{Factors} & \multicolumn{2}{c}{VCTK-IEMO} & \multicolumn{2}{c}{AISHELL-ESDB}  \\
   \cmidrule(r){2-4}  \cmidrule(r){5-6}  \cmidrule(r){7-8}
       & \multicolumn{1}{l}{Content} & \multicolumn{1}{l}{Rhythm} & \multicolumn{1}{l}{Pitch} & \multicolumn{1}{l}{IEMO} & \multicolumn{1}{l}{ESDB} & \multicolumn{1}{l}{IEMO} & \multicolumn{1}{l}{ESDB} \\
   \cmidrule(r){1-1}  \cmidrule(r){2-4}  \cmidrule(r){5-6}  \cmidrule(r){7-8}
   1  & -        & -        & -        & 59.08  & 28.08  & 30.39  & 80.58  \\
   2  & \cmark   & \cmark   & \cmark   & 58.46  & 27.37  & 30.35  & 75.46  \\
   3  & \xmark   & \xmark   & \xmark   & 25.00  & 24.83  & 25.00  & 25.69  \\
   \cmidrule(r){1-1}  \cmidrule(r){2-4}  \cmidrule(r){5-6}  \cmidrule(r){7-8}
   4  & \cmark   & \xmark   & \xmark   & 43.27  & 24.96  & 25.00  & 43.62  \\
   5  & \xmark   & \cmark   & \xmark   & 57.85  & 29.88  & 21.92  & 72.67  \\
   6  & \xmark   & \xmark   & \cmark   & 41.74  & 21.62  & 25.00  & 36.54  \\
   \cmidrule(r){1-1}  \cmidrule(r){2-4}  \cmidrule(r){5-6}  \cmidrule(r){7-8}
   7  & \cmark   & \cmark   & \xmark   & 57.14  & 30.33  & 19.27  & 73.08  \\
   8  & \cmark   & \xmark   & \cmark   & 40.20  & 25.00  & 24.77  & 44.96  \\
   9  & \xmark   & \cmark   & \cmark   & 57.89  & 29.33  & 21.66  & 71.58  \\
   \bottomrule
  \end{tabular}}
\end{table}

Firstly, it can be observed that although the results of the within-lingual tests are highly promising, the performance of the cross-lingual tests is very bad, sometimes even worse than guess (25\%).
Besides, by removing one or several unimportant components, the cross-lingual performance can be increased rather than decreased (ref. to System 2 and System 5, 7, 9, with IEMO training and 
ESDB test). This demonstrates the potential of the decomposition approach towards a generalizable emotion recognition,
and also suggests that information selection/control might be important for cross-lingual SER.

Nevertheless, since all the results are so bad in the cross-lingual tests, no conclusions can be said convincing. Perhaps the only thing we can make sure is that the present 
DNN-based SER system is not well generalized and more research is required.

\section{Conclusions}
\label{sec:con}

In this paper, we presented a preliminary study on how DNN-based SER models make decisions on emotions.
To answer this question, we employed the SpeechFlow model to decompose speech signals into separate information factors,
and then comprehensively studied the impact of each information factor and their combinations to the SER performance.
Our results on three emotion corpora showed that rhythm has more discrimination and generalization capability on SER,
at least in within-corpus tests.
However, with more challenging mismatch such as in the cross-lingual test, all these factors and their combinations failed to 
deliver reasonable performance.
This demonstrated that the current speech emotion model is still unreliable and can not be applied `in the wild'.
As for the future work, we will study more powerful factorization models, in order to pursue more independent factors. Moreover, 
we need larger emotion datasets to verify the observations in this study. Finally, the impact of cultural discrepancy deserves deeper investigation,
by which we may explain the weak performance in our cross-lingual test.



\bibliographystyle{IEEEtran}
\bibliography{refs}

\end{document}